\documentclass[11pt]{revtex4}    
\usepackage{amssymb,epsf}
\usepackage{latexsym}
\begin{document}
\title{Topological Born-Infeld-dilaton black holes}
\author{Ahmad Sheykhi \footnote{sheykhi@mail.uk.ac.ir}}
\address{Department of Physics, Shahid Bahonar University, P.O. Box 76175-132, Kerman, Iran\\
Research Institute for Astronomy and Astrophysics of Maragha
(RIAAM), P.O. Box 55134-441, Maragha, Iran}

\begin{abstract}
We construct a new analytic solution of
Einstein-Born-Infeld-dilaton theory in the presence of
Liouville-type potentials for the dilaton field. These solutions
describe dilaton black holes with nontrivial topology and
nonlinear electrodynamics. Black hole horizons and cosmological
horizons in these spacetimes, can be a two-dimensional positive,
zero or negative constant curvature surface. The asymptotic
behavior of these solutions are neither flat nor (A)dS. We
calculate the conserved and thermodynamic quantities of these
solutions and verify that these quantities satisfy the first law
of black hole thermodynamics.

\end{abstract}

 \maketitle

\section{Introduction}

Although the nonlinear electrodynamics was first introduced
several decades ago by Born and Infeld for the purpose of solving
various problems of divergence appearing in the Maxwell theory
\cite{BI}, in recent years, the study of the nonlinear
electrodynamics got a new impetus. Strong motivation comes from
developments in string/M-theory, which is a promising approach to
quantum gravity. It has been shown that the Born-Infeld theory
naturally arises in the low energy limit of the open string theory
\cite{Frad,Cal}. The Born-Infeld action including a dilaton and an
axion field, appears in the coupling of an open superstring and an
Abelian gauge field theory \cite{Frad}. This action, describing a
Born-Infeld-dilaton-axion system coupled to Einstein gravity, can
be considered as a nonlinear extension in the Abelian field of
Einstein-Maxwell-dilaton-axion gravity. Although one can
consistently truncate such models, the presence of the dilaton
field cannot be ignored if one considers coupling of the gravity
to other gauge fields, and therefore one remains with
Einstein-Born-Infeld gravity in the presence of a dilaton field.
Many attempts have been done to construct solutions of
Einstein-Born-Infeld-dilaton (EBId) gravity
\cite{Tam1,Tam2,YI,yaz, Clement,yazad,SRM,Shey,DHSR}. The
appearance of the dilaton field changes the asymptotic behavior of
the solutions to be neither asymptotically flat nor (A)dS. The
motivation for studying non asymptotically flat nor (A)dS
solutions of Einstein gravity comes from the fact that, these kind
of solutions can shed some light on the possible extensions of
AdS/CFT correspondence. Indeed, it has been speculated that the
linear dilaton spacetimes, which arise as near-horizon limits of
dilatonic black holes, might exhibit holography \cite{Ahar}. Black
hole spacetimes which are neither asymptotically flat nor (A)dS
have been explored widely in the literature
\cite{MW,PW,CHM,Clem,Mitra,Shey0, SR,DF,Dehmag,SDR,SDRP,yaz2}. In
the absence of dilaton field, exact solutions of
Einstein-Born-Infeld theory with/without cosmological constant
have  been constructed in
\cite{Gar,Wil,Aie,Tamaki,Fern,Dey,Cai1,Deh3}. In the scalar-tensor
theories of gravity, black hole solutions coupled to a Born-Infeld
nonlinear electrodynamics have also been studied recently in
\cite{yaz3}.

On the other hand, it is a general belief that in four dimensions
the topology of the event horizon of an asymptotically flat
stationary black hole is uniquely determined to be the two-sphere
$S^2$ \cite{Haw1,Haw2}. Hawking's theorem requires the integrated
Ricci scalar curvature with respect to the induced metric on the
event horizon to be positive \cite{Haw1}. This condition applied
to two-dimensional manifolds determines uniquely the topology. The
``topological censorship theorem" of Friedmann, Schleich and Witt
is another indication of the impossibility of non spherical
horizons \cite{FSW1,FSW2}. However, when the asymptotic flatness
of spacetime is violated, there is no fundamental reason to forbid
the existence of static or stationary black holes with nontrivial
topologies. It has been shown that for asymptotically AdS
spacetime, in the four-dimensional Einstein-Maxwell theory, there
exist black hole solutions whose event horizons may have zero or
negative constant curvature and their topologies are no longer the
two-sphere $S^2$. The properties of these black holes are quite
different from those of black holes with usual spherical topology
horizon, due to the different topological structures of the event
horizons. Besides, the black hole thermodynamics is drastically
affected by the topology of the event horizon. It was argued that
the Hawking-Page phase transition \cite{Haw3} for the
Schwarzschild-AdS black hole does not occur for locally AdS black
holes whose horizons have vanishing or negative constant
curvature, and they are thermally stable \cite{Birm}. The studies
on the topological black holes have been carried out extensively
in many aspects \cite{Lemos,Cai2,Bril1,Cai3,Cai4,
Shey1,Cri,MHD,other,Ban}. In this Letter we will construct a new
analytic solutions in four dimensional EBId theory. These
solutions which describe dilaton black holes with nonlinear
electrodynamics, and nontrivial topology, have unusual
asymptotics. They are neither asymptotically flat nor (A)dS. We
compute the conserved quantities of these solutions and find out
that they satisfy the first law of black hole thermodynamics.

\section{Basic Equations and Solutions}\label{Field}

We examine the action in which gravity is coupled to dilaton and
Born-Infeld fields
\begin{equation}\label{Act}
S=\frac{1}{16\pi}\int{d^{4}x\sqrt{-g}\left(\mathcal{R}\text{ }-2
g^{\mu\nu}
\partial_{\mu} \Phi \partial_{\nu}\Phi -V(\Phi
)+L(F,\Phi)\right)},
\end{equation}
where $\mathcal{R}$ is the Ricci scalar curvature, $\Phi $ is the
dilaton field and $V(\Phi )$ is a potential for $\Phi $. The
Born-Infeld $L(F,\Phi)$ part of the action is given by
\begin{equation}
L(F,\Phi)=4\beta^{2} e^{2\alpha \Phi}\left( 1-\sqrt{1+\frac{e^{-
4\alpha \Phi}F^2}{2\beta^{2}}}\right).
\end{equation}
Here, $\alpha $ is a constant determining the strength of coupling
of the scalar and electromagnetic field  and $F^2=F_{\mu \nu
}F^{\mu \nu }$, where $F_{\mu \nu }$ is the electromagnetic field
tensor.  $\beta $ is called the Born-Infeld parameter with
dimension of mass. It is worth noting that we have adopted,
following \cite{Tam1}, the open string version of the Born-Infeld
action coupled to a dilaton field since we would like to examine
the pure electric case. Clearly, this version of the
Born-Infeld-dilaton action does not enjoy electric-magnetic
duality \cite{Clement}. This form for the Born-Infeld-dilaton term
have been investigated previously by a number of authors (see e.g.
\cite{Tam1,Tam2,YI,yaz}). In the limit $\beta \rightarrow \infty
$, $L(F,\Phi)$ reduces to the standard Maxwell field coupled to a
dilaton field
\begin{equation}
L(F,\Phi)=-e^{-2\alpha \Phi }F_{\mu \nu }F^{\mu \nu }.
\end{equation}
On the other hand, $L(F,\Phi)\rightarrow 0$ as $\beta \rightarrow
0$. It is convenient to set
\begin{equation}
L(F,\Phi)=4\beta^2 e^{2\alpha \Phi}{\mathcal{L}}(Y),
\end{equation}
where
\begin{eqnarray}
{\mathcal{L}}(Y) &=&1-\sqrt{1+Y},\label{LY}\\
Y&=& \frac{e^{-4\alpha \Phi}F^2}{2\beta^2}.\label{Y}
\end{eqnarray}
By varying the action (\ref{Act}) with respect to the
gravitational field $g_{\mu \nu }$, the dilaton field $\Phi $ and
the electromagnetic field $A_{\mu }$ we obtain the equations of
motion
\begin{eqnarray}\label{FE1}
{\cal R}_{\mu\nu}&=& 2 \partial _{\mu }\Phi
\partial _{\nu }\Phi +\frac{1}{2}g_{\mu \nu }V(\Phi)-
4e^{-2\alpha \Phi}\partial_{Y}{{\cal L}}(Y) F_{\mu\eta}
F_{\nu}^{\text{ }\eta } \nonumber \\
&&+2\beta^2 e^{2\alpha \Phi}
\left[2Y\partial_{Y}{{\cal L}}(Y)-{{\cal L}}(Y)\right]g_{\mu\nu},
\end{eqnarray}
\begin{equation}\label{FE2}
\nabla ^{2}\Phi =\frac{1}{4}\frac{\partial V}{\partial \Phi}+
2\alpha \beta^2 e^{2\alpha \Phi }\left[2{ Y}\partial_{Y}{{\cal
L}}(Y)-{\cal L}(Y)\right],
\end{equation}
\begin{equation}\label{FE3}
\nabla _{\mu }\left(e^{-2\alpha \Phi}
\partial_{Y}{{\cal L}}(Y) F^{\mu\nu}\right)=0.
\end{equation}
Note that in the case of the linear electrodynamics with ${\cal
L}(Y)=-{1\over 2}Y$, the system of equations
(\ref{FE1})-(\ref{FE3}) reduce to the well-known equations of
Einstein-Maxwell-dilaton (EMd) gravity \cite{CHM}.

We would like to find topological solutions of the above field
equations. The most general such metric can be written in the form
\begin{equation}\label{metric}
ds^2=-f(r)dt^2 +{dr^2\over f(r)}+ r^2R^2(r)d\Omega_{k}^2 ,
\end{equation}
where $f(r)$ and $R(r)$ are functions of $r$ which should be
determined, and $d\Omega_{k}^2$ is the line element of a
two-dimensional hypersurface $\Sigma$ with constant curvature,
\begin{equation}\label{met}
d\Omega_k^2=\left\{
  \begin{array}{ll}
    $$d\theta^2+\sin^2\theta d\phi^2$$,\quad \quad\!\!{\rm for}\quad $$k=1$$, &  \\
    $$d\theta^2+\theta^2 d\phi^2$$,\quad\quad\quad {\rm for}\quad $$k=0$$,&  \\
    $$d\theta^2+\sinh^2\theta d\phi^2$$, \quad {\rm for}\quad $$k=-1$$.&
  \end{array}
\right.
\end{equation}
For $k = 1$, the topology of the event horizon is the two-sphere
$S^2$, and the spacetime has the topology $R^2 \times S^2$. For $k
= 0$, the topology of the event horizon is that of a torus and the
spacetime has the topology $R^2 \times T^2$. For $k = -1$, the
surface $\Sigma$ is a $2$-dimensional hypersurface $H^2$ with
constant negative curvature. In this case the topology of
spacetime is $R^2 \times H^2$. First of all, the electromagnetic
fields equation (\ref{FE3}) can be integrated immediately, where
all the components of $F_{\mu\nu}$ are zero except $ F_{tr}$:
\begin{equation}\label{Ftr}
F_{tr}=\frac{\beta q e^{2\alpha \Phi}}{\sqrt{\beta^2 \left(
rR\right)^{4}+q^{2}}},
\end{equation}
where $q$, is an integration constant related to the electric
charge of the black hole. Defining the electric charge via  $ Q =
\frac{1}{4\pi} \int e^{-2\alpha\Phi}\text{ }^{*} F d{\Omega}, $ we
get
\begin{equation}
{Q}=\frac{q\omega}{4\pi},  \label{Charge}
\end{equation}
where $\omega$ represents the area of the constant hypersurface
$\Sigma$. It is worthwhile to note that the electric field is
finite at $r=0$. This is expected in Born-Infeld theories.
Meanwhile it is interesting to consider three limits of Eq.
(\ref{Ftr}). First, for large $\beta$ (where the BI action reduces
to Maxwell case) we have $F_{tr}=q e^{2\alpha \Phi}/(rR)^{2}$ as
presented in \cite{CHM}. On the other hand, if $\beta\rightarrow
0$ we get $F_{tr}=0$. Finally, in the absence of the dilaton field
($\alpha=0$), it reduces to the case of Einstein-Born-Infeld
theory \cite{Dey}
\begin{equation}
F_{tr}=\frac{ \beta q }{\sqrt{\beta^2 r^4+q^{2}}}.
\end{equation}
Our aim here is to construct exact topological solutions of the
EBId theory with an arbitrary dilaton coupling constant $\alpha$.
The case in which we find topological solutions of physically
interest is to take the dilaton potential of the form
\begin{equation}\label{v2}
V(\Phi) = 2\Lambda_{0} e^{2\zeta_{0}\Phi} +2 \Lambda e^{2\zeta
\Phi},
\end{equation}
where $\Lambda_{0}$,  $\Lambda$, $ \zeta_{0}$ and $ \zeta$ are
constants. This kind of  potential was previously investigated by
a number of authors both in the context of
Friedman-Robertson-Walker (FRW) scalar field cosmologies
\cite{ozer} and EMd black holes (see e.g
\cite{CHM,yaz2,SRM,Shey}). In order to solve the system of
equations (\ref{FE1}) and (\ref{FE2}) for three unknown functions
$f(r)$, $R(r)$ and $\Phi (r)$, we make the ansatz \cite{DF}
\begin{equation}
R(r)=e^{\alpha \Phi}.\label{Rphi}
\end{equation}
A motivation for taking this ansatz is that in the absence of a
dilaton field ($\alpha=0$) it reduces to $R(r)=1$, as one expected
(see Eq. \ref{metric}). Another motivation comes from the form of
the electromagnetic field equation provided one write it
explicitly in terms of partial derivative \cite{DF}. Inserting
(\ref{Rphi}), the electromagnetic field (\ref{Ftr}) and the metric
(\ref{metric}) into the field equations (\ref{FE1}) and
(\ref{FE2}), one can show that these equations have the following
solutions
\begin{eqnarray}
f(r)&=&-k\frac{ {\alpha}^{2}+1}{ {\alpha}^{2}-1
}{b}^{-2\gamma}{r}^{2\gamma}-\frac{m}{r^{1-2\gamma}}+\frac{(\Lambda-2\beta^2)
\left( {\alpha}^{2}+1 \right) ^{2}{b}^{2 \gamma}}{\alpha^{2}-3
}r^{2-2\gamma}\nonumber \\
&&-2\beta ^{2}\left( \alpha ^{2}+1\right) b^{2\gamma }r^{2\gamma
-1}\int r ^{2(1-2\gamma)}\sqrt{ 1+\eta}{dr }, \nonumber
\\ \label{f1}
\end{eqnarray}
\begin{equation}\label{phi}
\Phi (r)=\frac{\alpha }{\alpha ^{2}+1}\ln (\frac{b}{r}),
\end{equation}
where $b$ is an arbitrary constant, $\gamma =\alpha ^{2}/(1+\alpha
^{2})$, and
\begin{equation}
\eta \equiv \frac{q^{2}}{\beta ^{2}b^{4\gamma }r^{4(1-\gamma )}}.
\label{eta}
\end{equation}
In the above expression, $m$ appears as an integration constant
and is related to the ADM (Arnowitt-Deser-Misner) mass of the
black hole. According to the definition of mass due to Abbott and
Deser \cite{abot}, the mass of the solution (\ref{f1}) is
\begin{equation}
{M}=\frac{b^{2\gamma}m\omega }{8\pi(\alpha^2+1)}. \label{Mass}
\end{equation}
The above solutions will fully satisfy the system of equations
(\ref{FE1}) and (\ref{FE2}) provided we have $\zeta_{0}
=1/\alpha$,  $\zeta=\alpha$ and $ \Lambda_{0}= {k b^{-2}\alpha^2
}/{(\alpha^2-1)}. $ Notice that $\Lambda$ remains as a free
parameter which plays the role of the cosmological constant. For
later convenience, we redefine it as $\Lambda=-3/l^2$, where $l$
is a constant with dimension of length. One may refer to $\Lambda$
as the cosmological constant, since in the absence of the dilaton
field ($\alpha=0$) the action reduces to the action of EBI gravity
with cosmological constant \cite{Dey,Cai1}. The integral can be
done in terms of hypergeometric function and can be written in a
compact form. The result is
\begin{eqnarray}\label{f2}
f(r) &=&-k\frac{ {\alpha}^{2}+1}{ {\alpha}^{2}-1
}{b}^{-2\gamma}{r}^{2\gamma}-\frac{m}{r^{1-2\gamma}}
+\frac{\Lambda\left( {\alpha}^{2}+1 \right) ^{2}{b}^{2
\gamma}}{\alpha^{2}-3}r^{2-2\gamma}\nonumber \\
&&-\frac{2\beta^2\left( {\alpha}^{2}+1 \right) ^{2}{b}^{2
\gamma}}{\alpha^{2}-3
}r^{2-2\gamma}\times \left( 1-\text{{\ }}%
_{2}F_{1}\left( -\frac{1}{2},\frac{\alpha ^{2}-3}{4} ,
\frac{\alpha ^{2}+1}{4} ,-\eta \right) \right).
\end{eqnarray}
It is worthwhile to compare our solution to those presented in
\cite{Clement}. The authors of \cite{Clement} investigated
asymptotically flat static and spherically symmetric solutions in
Born-Infeld-dilaton theory in the absence of the dilaton
potential. They adopted the $SL(2,R)$ symmetry version of the
Born-Infeld-dilaton action. Imposing the asymptotically flat
condition, they found a class of black hole solutions in terms of
series expansion and explored their properties. Our solutions are
different from their solutions for several reasons. First, the
coupling of the dilaton to the Born-Infeld term in the action, for
which there is no symmetry between electric and magnetic
solutions, in contrast to the action of \cite{Clement} which has
$SL(2,R)$ $S$-dual symmetry. Second, the presence of the
Liouville-type potentials (the negative effective cosmological
constant) in our case, which play a crucial role in the existence
of these topological dilaton black holes. And third the asymptotic
behaviour of the solution which changes to be neither flat nor
(A)dS due to the presence of the dilaton potential. One may note
that as $\beta \longrightarrow \infty $ these solutions reduce to
the topological black hole solutions in EMd gravity given in Ref.
\cite{Shey1}. In the absence of a dilaton field ($\alpha =\gamma
=0 $), the above solutions reduce to
\begin{eqnarray}
f(r) &=&k-\frac{m}{r}+\frac{r^2}{l^2}
+\frac{2\beta^2 }{3}r^2 \times  \left( 1-\text{{\ }}%
_{2}F_{1}\left( -\frac{1}{2},-\frac{3}{4},
\frac{1}{4},-\frac{q^{2}}{\beta ^{2}r^4} \right) \right),
\end{eqnarray}
which describes an asymptotically AdS topological Born-Infeld
black hole with positive, zero or negative constant curvature
hypersurface \cite{Cai1}. Using the fact that $_2F_1(a,b,c,z)$ has
a convergent series expansion for $|z| <1$, we can find the
behavior of (\ref{f2}) for large $r$. This is given by
\begin{eqnarray}
f(r) &=&-k\frac{ {\alpha}^{2}+1}{ {\alpha}^{2}-1
}{b}^{-2\gamma}{r}^{2\gamma}-\frac{m}{r^{1-2\gamma}}-\frac{3\left(
{\alpha}^{2}+1 \right) ^{2}{b}^{2 \gamma}}{l^2(\alpha^{2}-3)
}r^{2-2\gamma}\nonumber \\
&&+\frac{(\alpha^2+1)b^{-2\gamma}q^{2}}{r^{2-2\gamma}}-
\frac{(\alpha^{2}+1 )^{2}b^{-6\gamma }q^{4}}{4\beta^2(\alpha
^{2}+5)r^{6(1-\gamma )}}.
\end{eqnarray}
Note that for $\alpha =\gamma= 0$, the above expression reduces to
\begin{eqnarray}
f(r) &=&k-\frac{m}{r}+\frac{r^2}{l^2}+\frac{q^{2}}{r^{2}}-
\frac{1}{20\beta^2}\frac{q^{4}}{r^{6}},
\end{eqnarray}
which has the form of topological charged black hole in AdS
spacetime in the limit $\beta\rightarrow \infty$
\cite{Bril1}\cite{Cai3}. The last term in the right hand side of
the above expression is the leading Born-Infeld correction to the
topological black hole in the large $\beta$ limit.

\begin{figure}[tbp]
\epsfxsize=7cm \centerline{\epsffile{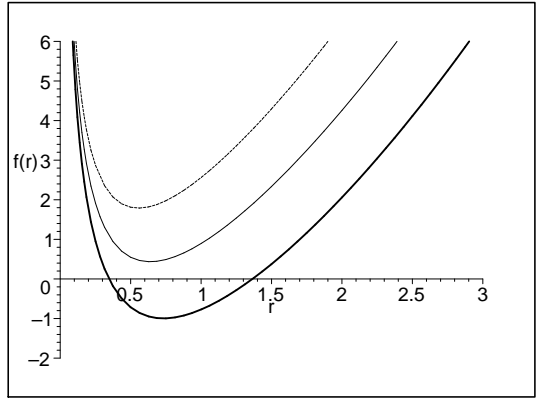}} \caption{The
function $f(r)$ versus $r$ for $\protect\alpha=0.5$, $m=2$,
$\beta=1$ and $q=1$. $k=-1$ (bold line), $k=0$ (continuous line)
and $k=1$ (dashed line).} \label{figure1}
\end{figure}

\begin{figure}[tbp]
\epsfxsize=7cm \centerline{\epsffile{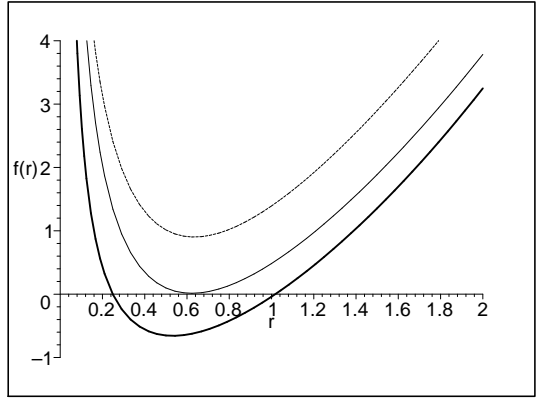}} \caption{The
function $f(r)$ versus $r$ for $m=2$, $\beta=1$, $q=1$ and $k=0$.
$\protect\alpha=0$ (bold line), $\protect\alpha=0.38$ (continuous
line) and $\protect\alpha=0.6$ (dashed line).} \label{figure2}
\end{figure}

\begin{figure}[tbp]
\epsfxsize=7cm \centerline{\epsffile{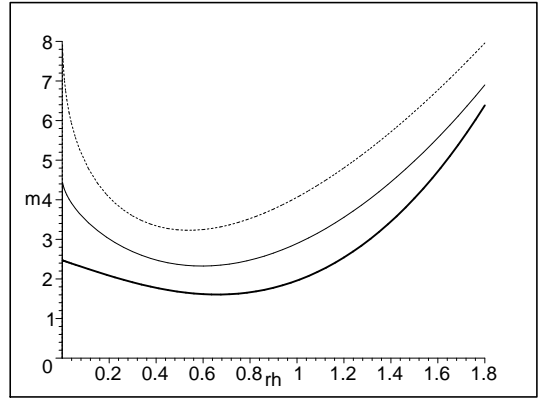}} \caption{The
function $m(r_h)$ versus $r_h$ for  $\beta=1$, $q=1$ and $k=0$.
$\protect\alpha=0$ (bold line), $\protect\alpha=0.5$ (continuous
line) and $\protect\alpha=0.7$ (dashed line).} \label{figure3}
\end{figure}

\begin{figure}[tbp]
\epsfxsize=7cm \centerline{\epsffile{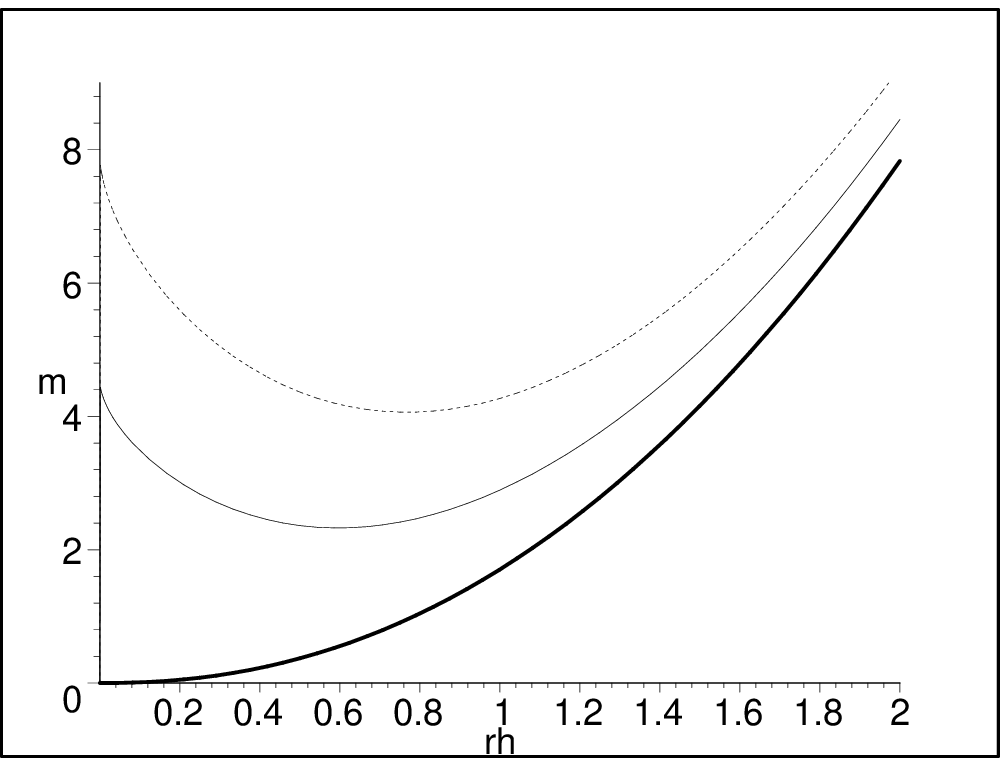}} \caption{The
function $m(r_h)$ versus $r_h$ for  $\beta=1$,
$\protect\alpha=0.5$ and $k=0$. $q=0$  (bold line), $q=1$
(continuous line) and $q=1.5$  (dashed line).} \label{figure4}
\end{figure}

\begin{figure}[tbp]
\epsfxsize=7cm \centerline{\epsffile{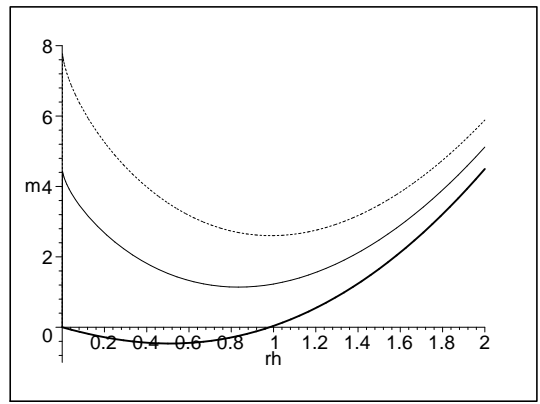}} \caption{The
function $m(r_h)$ versus $r_h$ for  $\beta=1$,
$\protect\alpha=0.5$ and $k=-1$. $q=0$  (bold line), $q=1$
(continuous line) and $q=1.5$  (dashed line).} \label{figure5}
\end{figure}
Next we study the physical properties of these solutions. For this
purpose, we first look for the curvature singularities. In the
presence of dilaton field, the Kretschmann scalar $R_{\mu \nu
\lambda \kappa }R^{\mu \nu \lambda \kappa }$ diverges at $r=0$, it
is finite for $r\neq 0$ and goes to zero as $r\rightarrow \infty
$. Thus, there is an essential singularity located at $r=0$. The
spacetime is neither asymptotically flat nor (A)dS. It is notable
to mention that in the $k=\pm1$ cases this solution does not exist
for the string case where $\alpha=1$. As one can see from Eq.
(\ref{f2}), the solution is ill-defined for $\alpha =\sqrt{3}$.
The cases with $\alpha <\sqrt{3}$ and $\alpha
>\sqrt{3}$ should be considered separately. In the first case
where $ \alpha <\sqrt{3}$,  there exist a cosmological horizon for
$\Lambda >0$, while there is no cosmological horizons if $\Lambda
<0$. In the latter case ($\alpha <\sqrt{3}$ and $\Lambda <0$) the
spacetimes associated with the solution (\ref{f2}) exhibit a
variety of possible casual structures depending on the values of
the metric parameters $\alpha $, $m$, $q$ and $k$ (see figs.
\ref{figure1}-\ref{figure2}). For simplicity in these figures, we
kept fixed the other parameters $l=b=1$. Figure \ref{figure1}
shows that for fixed value of other parameters, the number of
horizons increase with decreasing the constant curvature $k$,
while one can see from figure \ref{figure2} that with increasing
$\alpha$, the number of horizons decrease. In summary , these
figures show that our solutions can represent topological black
hole, with two horizons, an extreme topological black hole or a
naked singularity depending on the values of the metric
parameters. In the second case where $\alpha
>\sqrt{3}$, the spacetime has a cosmological horizon. Although, in principle, the casual structure of the spacetime can be obtained
by finding the roots of $ f(r)=0$, but because of the nature of
the dilaton and nonlinear electrodynamic fields in (\ref{f2}), it
is not possible to find analytically the location of the horizons.
To have further understanding on the nature of the horizons, we
plot in figures \ref{figure3}-\ref{figure5}, the mass parameter
$m$ as a function of the horizon radius $r_h$ for different value
of dilaton coupling constant $\alpha$, charge parameter $q$ and
curvature constant $k$. Again, we have fixed $l=b=1$, for
simplicity. It is easy to show that the mass parameter of the
black hole can be expressed in terms of the horizon radius $r_{h}$
as
\begin{eqnarray}\label{mass}
m(r_{h}) &=&-{\frac { k\left( { \alpha}^{2}+1 \right)
{b}^{-2\gamma}}{ { \alpha}^{2}-1  }}{r_{h}}+\frac{\Lambda \left(
{\alpha}^{2}+1 \right) ^{2}{b}^{2 \gamma}}{(\alpha^{2}-3
)}r_{h}^{3-4\gamma}\nonumber\\
&&-\frac{2\beta^2 (\alpha ^{2}+1)^{2}b^{2\gamma }}{(\alpha
^{2}-3)}r_{h}^{3-4\gamma} \times
\left(1-\text{}_{2}F_{1}\left( -\frac{1}{2},\frac{\alpha ^{2}-3%
}{4},\frac{\alpha ^{2}+1}{4} ,-\eta_{h}\right) \right).
\end{eqnarray}
where $\eta_{h}=\eta(r=r_{h})$. These figures show that for a
given value of $\alpha$, the number of horizons depend on the
choice of the value of the mass parameter $m$. We see that, up to
a certain value of the mass parameter $m$, there are two horizons,
and as we decrease the $m$ further, the two horizons meet. In this
case we get extremal black hole (see the next section). Figure
\ref{figure3} also shows that, as an example for $k=0$, with
increasing $\alpha$, the $m_{\mathrm{ext}}$ also increases. It is
worth noting that in the limit $r_{h}\rightarrow0$ we have a
nonzero value for the mass parameter $m$. This is in contrast to
the Schwarzschild black holes in which mass parameter goes to zero
as $r_{h}\rightarrow0$. As we have shown in figure \ref{figure4},
this is due to the effect of the charge parameter $q$ and the
nature of the Born-Infeld field, and in the case q=0, the mass
parameter $m$ goes to zero as $r_{h}\rightarrow0$. Besides in the
case $k=-1$ and $q=0$, our solution has strange properties. In
this case, one can see from figure \ref{figure5} that the mass of
the solution can be negative, however one still has a topological
black hole solution with negative curvature horizons. It was
argued that this kind of black hole with negative mass can also be
formed as a result of gravitational collapse \cite{Mann2}.


\section{Thermodynamics of topological black holes} \label{Therm}
In this section we would like to study the thermodynamical
properties of the topological dilaton black holes we have just
found. The Hawking temperature of the topological black hole on
the outer horizon $r_{+}$ can be calculated using the relation
\begin{equation}
T_{+}=\frac{\kappa}{2\pi}= \frac{f^{\text{ }^{\prime
}}(r_{+})}{4\pi},
\end{equation}
where $\kappa$ is the surface gravity. Then, one can easily show
that
\begin{eqnarray}\label{Tem}
T_{+}&=&-\frac{(\alpha ^2+1)}{4\pi
}b^{2\gamma}r_{+}^{1-2\gamma}\left(
\frac{kb^{-4\gamma}}{\alpha^2-1}r_{+}^{4\gamma-2}+\Lambda -2\beta^2(1-\sqrt{1+\eta_{+}})\right)\nonumber\\
&=&-\frac{kb^{-2\gamma}}{2\pi}r_{+}^{2\gamma-1}-\frac{(\alpha^{2}-3
)m}{4\pi(\alpha
^{2}+1)}{r_{+}}^{2\gamma-2}-\frac{q^2b^{-2\gamma}}{\pi}r_{+}^{2\gamma-3}\nonumber\\
&&\times \text{ }_{2}F_{1}\left(
\frac{1}{2},\frac{{\alpha}^{2}+1}{4} ,\frac{{\alpha }^{2}+5}{4}
,-\eta_{+} \right),
\end{eqnarray}
where $\eta_{+}=\eta(r=r_{+})$. The temperature of the black hole
is zero in the case of extremal black hole. It is easy to show
that
\begin{eqnarray}\label{mext}
m_{\mathrm{ext}}&=&-\frac{2k(\alpha ^2+1)b^{-2\gamma}}{\alpha
^{2}-3}r_{+}-\frac{4q^2(\alpha ^{2}+1)b^{-2\gamma}}{(\alpha
^{2}-3)r_{+}}\times \text{
}_{2}F_{1}\left(\frac{1}{2},\frac{{\alpha }^{2}+1}{4} ,
\frac{{\alpha }^{2}+5}{4} ,-\eta_{+}\right).
\end{eqnarray}
Indeed, the metric of Eqs. (\ref{metric}) and (\ref{f2}) can
describe a topological dilaton black hole with inner and outer
event horizons located at $r_{-}$ and $r_{+}$, provided
$m>m_{\mathrm{ext}}$, an extreme topological black hole in the
case of $m=m_{\mathrm{ext}}$, and a naked singularity if
$m<m_{\mathrm{ext}}$. It is worth noting that in the absence of a
nontrivial dilaton field ($\alpha=\gamma =0 $), expressions
(\ref{mass})-(\ref{mext}) reduce to that of an asymptotically AdS
topological black hole in Born-Infeld theory \cite{Cai1}.

\begin{figure}[tbp]
\epsfxsize=7cm \centerline{\epsffile{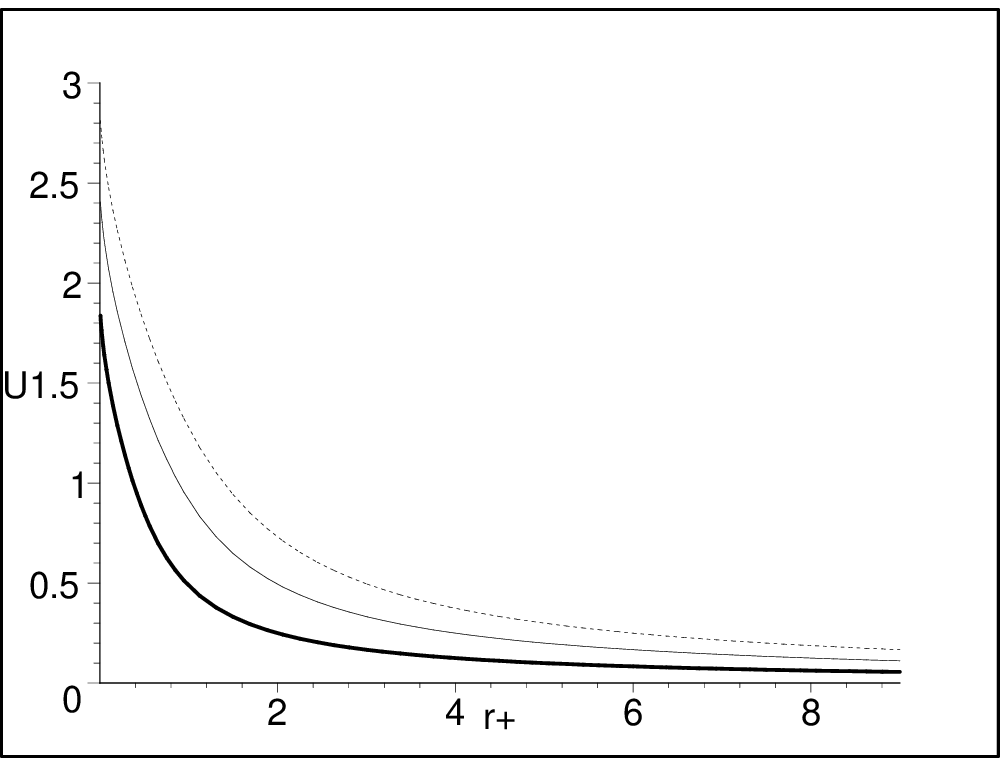}} \caption{The
function $U(r_+)$ versus $r_+$ for $\beta=1$, $b=1$ and
$\protect\alpha=0.5$.  $q=0.5$  (bold line), $q=1$ (continuous
line) and $q=1.5$  (dashed line).} \label{figure6}
\end{figure}

\begin{figure}[tbp]
\epsfxsize=7cm \centerline{\epsffile{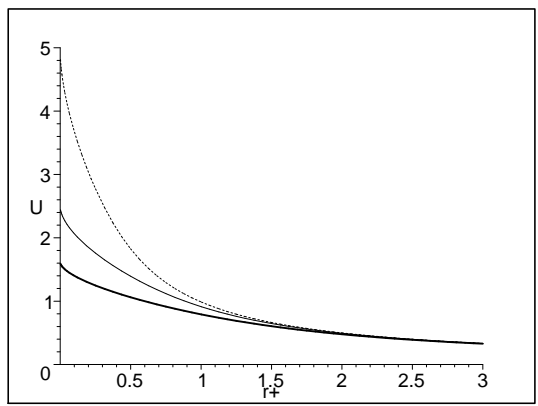}} \caption{The
function $U(r_+)$ versus $r_+$ for $q=1$, $b=1$ and
$\protect\alpha=0.5$.  $\beta=0.5$  (bold line), $\beta=1$
(continuous line) and $\beta=3$  (dashed line).} \label{figure7}
\end{figure}

The entropy of the topological black hole still obeys the so
called area law of the entropy which states that the entropy of
the black hole is a quarter of the event horizon area \cite{Beck}.
This near universal law applies to almost all kinds of black
holes, including dilaton black holes, in Einstein gravity
\cite{hunt}. It is a matter of calculation to show that the
entropy of the topological black hole is
\begin{equation}
{S}=\frac{b^{2\gamma}r_{+}^{2(1-\gamma
)}\omega}{4}.\label{Entropy}
\end{equation}
The electric potential $U$, measured at infinity with respect to
the horizon, is defined by
\begin{equation}
U=A_{\mu }\chi ^{\mu }\left| _{r\rightarrow \infty }-A_{\mu }\chi
^{\mu }\right| _{r=r_{+}},  \label{Pot}
\end{equation}
where $\chi=\partial_{t}$ is the null generator of the horizon.
One can easily show that the gauge potential $A_{t }$
corresponding to the electromagnetic field (\ref{Ftr}) can be
written as
\begin{eqnarray}\label{vectorpot}
A_{t}&=&\frac{q}{r}\times \text{ }_{2}F_{1}\left(
\frac{1}{2},\frac{\alpha ^{2}+1}{4} , \frac{\alpha^{2}+5}{4}
,-\eta\right).
\end{eqnarray}
Therefore the electric potential may be obtained as
\begin{equation}
U=\frac{q}{r_{+}}\times \text{ }_{2}F_{1}\left(
\frac{1}{2},\frac{\alpha ^{2}+1}{4} , \frac{\alpha^{2}+5}{4}
,-\eta_{+}\right). \label{Pot}
\end{equation}
In figures \ref{figure6} and \ref{figure7} we have shown the
behavior of electric potential $U$ as a function of horizon
radius. As one can see from these figures, $U$ is finite even for
$r_+=0$.

Then, we consider the first law of thermodynamics for the
topological black hole. In order to do this, we obtain the mass
$M$ as a function of extensive quantities $S$ and $Q$. Using the
expression for the charge, the mass and the entropy given in Eqs.
(\ref{Charge}), (\ref{Mass}) and (\ref{Entropy}) and the fact that
$f(r_{+})=0$, one can obtain a Smarr-type formula as
\begin{eqnarray}
M(S,Q)&=&-\frac{k b^{-\alpha^2}{\left(4S\right)}^{(\alpha^2+1)/2}
}{8\pi(\alpha^2-1)} -\frac{3(\alpha^2+1)b^{\alpha^2}}{8\pi
l^2(\alpha^2-3)} {\left(4S\right)}^{(3-\alpha^2)/2}
\nonumber\\
&&-\frac{\beta^2(\alpha^2+1)b^{\alpha^2}}
{4\pi(\alpha^2-3)}{\left(4S\right)}^{(3-\alpha^2)/2}\times
\left(1-\text{ }_{2}F_{1}\left( -\frac{1}{2},\frac{\alpha ^{2}-3%
}{4},\frac{\alpha ^{2}+1}{4},\frac{-\pi^2Q^{2}}{%
\beta^2 S^2}\right)\right).\nonumber\\
\label{Msmar}
\end{eqnarray}
We can regard the parameters $S$, and $Q$ as a complete set of
extensive parameters for the mass $M(S,Q)$ and define the
intensive parameters conjugate to $S$ and $Q$. These quantities
are the temperature and the electric potential
\begin{equation}
T=\left( \frac{\partial M}{\partial S}\right) _{Q},\ \ U=\left( \frac{\partial M%
}{\partial Q}\right) _{S}.  \label{Dsmar}
\end{equation}
Numerical calculations show that the intensive quantities
calculated by Eq. (\ref{Dsmar}) coincide with Eqs. (\ref{Tem}) and
(\ref{Pot}). Thus, these thermodynamic quantities satisfy the
first law of black hole thermodynamics
\begin{equation}
dM = TdS+Ud{Q}.
\end{equation}
\section{Conclusions}
Black holes in AdS spacetimes are quite different from their
counterparts in asymptotically flat spacetime. In AdS spacetimes,
there are a kind of black holes with nontrivial topology. The
event horizons of these black holes can be a positive, zero or
negative constant curvature surface. These black holes are
generally called topological black holes. The construction and
analysis of these exotic black holes in AdS space is a subject of
much recent interest. This is primarily due to their relevance for
the AdS/CFT correspondence. In this letter, we further generalized
these exotic black hole solutions by including a dilaton and
nonlinear electrodynamic fields in the action. In contrast to the
topological black holes in the Einstein-Maxwell theory, which are
asymptotically AdS, the topological dilaton black holes we found
here, are neither asymptotically flat nor (A)dS. Indeed, the
Liouville-type potentials (the negative effective cosmological
constant) plays a crucial role in the existence of these black
hole solutions, as the negative cosmological constant does in the
Einstein-Maxwell theory. These solutions do not exist for the
string case where $\alpha=1$ provided $k=\pm1$. Besides they are
ill-defined for $\alpha \neq \sqrt{3}$. In the absence of a
dilaton field ($\alpha =\gamma=0 $), our solutions reduce to the
four-dimensional topological black hole solutions of Born-Infeld
theory \cite{Cai1}, while in the limit $\beta\rightarrow \infty$
they reduce to the topological black holes in
Einstein-Maxwell-dilaton gravity \cite{Shey1} (see also
\cite{Cai4}). We showed that our solutions can describe
topological Born-Infeld black hole with inner and outer event
horizons, an extreme topological black hole or a naked singularity
provided the parameters of the solutions are chosen suitably. We
also computed the charge, mass, temperature, entropy and electric
potential of the topological dilaton black holes and verified that
these quantities satisfy the first law of black hole
thermodynamics.

Although, in this Letter we constructed the topological
Born-Infeld-dilaton black holes in the presence of Liouville-type
potentials for the dilaton field, and discussed their
thermodynamical properties, many issues however still remain to be
investigated. We know that Reissner-Nordstrom AdS black holes
undergo Hawking-Page phase transition. This transition gets
modified as we include a dilaton and nonlinear electrodynamics
fields corrections into account. Indeed, the dilaton field, as
well as the Born-Infeld field, can create an unstable phase for
the solutions \cite{Shey}. A detail study on this issue for the
case of topological dilaton black holes in the presence of a
nonlinear electrodynamics will be addressed elsewhere
\cite{Shey2}. It would be also of great interest to generalize
these solutions to higher dimensional Einstein-Born-Infeld-dilaton
gravity \cite{Shey2}.

\acknowledgments{This work has been supported financially by
Research Institute for Astronomy and Astrophysics of Maragha,
Iran.}

\end{document}